# DIGITAL LITERACY AND READING HABITS OF THE DMI-ST. EUGENE UNIVERSITY STUDENTS

*Subaveerapandiyan*
*Priyanka Sinha*

**Mr. Subaveerapandiyan A.**
Librarian
DMI-St. Eugene University,
Lusaka, Zambia
Email:
subaveerapandiyan@gmail.com
**(Corresponding Author)**

**Ms. Priyanka Sinha**
Ph.D. Research Scholar
Department of Library and
Information Science
Punjab University, Chandigarh
Email:
priyankasinha101099@gmail.com

Digital literacy is the skill of finding, evaluating, consuming, and generating information using digital technologies. The study attempted to comprehend university students' digital reading habits and skills. It also provides a glimpse of the pupils' favorite reading materials, including physical and digital sources. We examined BSc and BE Computer Science students of DMI-St. Eugene University, Zambia. The tool was a structured questionnaire that was distributed through WhatsApp. The study's findings revealed that most students thoroughly understand digital tools and how to use them but lack the skills to build their websites and portfolio. Out of 115 students, all agreed they used computers for learning purposes. Usage of digital environments, generally, they used the World Wide Web for searching for information. Additionally, most students have medium digital application skills, despite their preference for reading electronic books. The results indicate that students' gender and level of education had a statistically significant link with their digital literacy, whereas age wasn't shown to be a statistically relevant predictor. The findings show that, in terms of education, especially reading, students' or readers' top priorities are electronic resources; print book preferences are reduced.

## INTRODUCTION

The term "digital literacy" is not new. Gilster coined the phrase "digital literacy" in 1997 to describe the ability to gather meaningful knowledge from various sources via the Internet. Digital literacy is the capacity to comprehend information and, more crucially, to assess and integrate information presented in multiple formats by a computer. Information evaluation and interpretation skills are essential (Pool, 1997). The capacity to use modern tools, devices, and digital technologies such as computers, phones, tablets, computer programs, and online media in the most effective, current, and efficient manner is known as digital literacy (DL). The ability to acquire, manage, integrate, analyze, and evaluate information, develop new knowledge, create, and interact with others utilizing digital technology and communication tools is called digital literacy (Potter, 2019).

The educational system has changed significantly since it was just applied as a traditional classroom tool to create online teaching and learning models. Digital literacy is very important for students as a basis for being able to compete in the world. The present generation is more likely to be digital natives as they are born in this technology age. However, a child's family should serve as the foundation for their education in digital literacy. Thus parents' awareness of the topic and adoption of good habits are essential. Marc Prensky coined the phrase "digital natives" in 2001 to refer to the pupils as the first generation who "spent their entire childhood surrounded by and using" digital technologies. He said, "The digital language of computers, video games, and the internet is what all our pupils today are 'native speakers of (Prensky, 2001)."

The Chartered Institute of Library defines information literacy, and Information Professionals (CILIP) definition gives us the terminology to start recognizing the overlap and logical progression of information literacy to digital literacy in library instruction: Everybody requires some specific set of skills and abilities, such as the ability to find, access, interpret, analyze, manage, produce, communicate, save, and share information (CILIP, 2018). A global framework to measure digital literacy was developed by UNESCO in 2018, and its pillars include information and data literacy, communication and cooperation, the creation of digital content, safety, problem-solving, and career-related skills. This approach was also used for the European Digital Competence Framework 2.0 inside the European Digital Agenda (*UNESCO*, 2018).

## REVIEW OF LITERATURE

### Digital Literacy

The notion of digital literacy has changed through time due to technological advancements, the growth of the web, a rise in Internet usage, and the incorporation of digital technology into various industries and aspects of people's life (Özparlak, 2022). Paul Glister coined the phrase "digital literacy" in 1997 to describe a person's capacity to comprehend and utilize information supplied through computers in various formats and from a wide range of sources (Reddy et al., 2020).

Digital literacy is the capacity to use digital technology to safely and ethically acquire, manage, comprehend, integrate, communicate, evaluate, and generate information for employment, decent employment, and entrepreneurship. It encompasses skills referred to variably as media literacy, information literacy, computer literacy, and ICT literacy (Teng, 2018). Digital literacy also refers to the variety of ways that people use digital texts and technologies to cooperate, create, and communicate. Without a critical perspective on online and offline power relations, we may also be constrained in these endeavors. Additionally, their social settings impact how they interact, produce and communicate (Manderino & Castek, 2020). In order to identify, access, manage, integrate, evaluate, analyze, and synthesize digital content, create new knowledge, and interact with others, a person must have digital literacy (LIBRe Foundation, 2016). Further, a four-gear model developed by (Reddy et al., 2020) illustrates how digital literacy is the key to any country's sustainable growth. Therefore, promoting digital literacy is essential for having a more increased sense of civic engagement, enhancing a nation's education system, and achieving sustainable development goals. One must have digital literacy to use digital technologies to create, curate, and consume knowledge.

Some studies mainly focus on university students. In one study, Sivrikaya (2020) evaluates the students' degrees of digital literacy at the school of sports sciences. 394 Atatürk University students participated in the study as respondents. The results show that the participating students' levels of digital literacy vary significantly in the social sub-dimension concerning gender variation. Male pupils had higher levels of digital literacy than female students in the social sub-dimension of the scale. In further studies, Zulkarnain et al. (2020) conducted a research survey to describe students' e-learning-based digital mathematics competence in the Covid-19 epoch. At Riau University the results of this study demonstrate the significance of digital literacy abilities that positively affect knowledge, comprehension, and skills in using media, particularly social media, which the general public and students increasingly frequently utilized as a source of information.

Göldag (2021) addressed the correlation between university students' degrees in digital skills and mindfulness of digital cybersecurity. The research's core population is made up of Inonu University students. Two hundred sixty-five pupils willingly participated. The findings show that students have a reasonable level of digital literacy and a high level of awareness of digital data protection. Male students had greater levels of digital literacy and understanding of data security than female pupils. Those with computers have greater digital literacy and understanding of digital data protection than students without computers. Pupils who use computers for extended periods every day have greater levels of technological proficiency and consciousness of digital data security than learners who would use computers for short times every day. Digital literacy and comprehension of concerns regarding data security among students rise as they utilize digital gadgets. The degree of pupils' digital literacy and understanding of digital data security are highly correlated and significant. Nguyen and Habók (2022) examined the digital literacy levels among 1661 English as a foreign language student at Vietnamese universities. They evaluated students' perceived digital skills, perspectives towards using digital tools, and frequency of using technological applications for learning English. The results show that most pupils have access to digital gadgets at home and in their places of education. Additionally, pupils learn enough about digital literacy, and their technological aptitudes range from below average to average. According to comparisons, males outperform their female counterparts in technological expertise and proficiency. In conclusion, digital literacy is a large and complicated subject, and mastering its concepts and skills is essential for academic achievement. Students in the twenty-first century may be digital natives.

**Reading Habits**

Reading is the most efficient approach for a person to develop his critical thinking abilities, gain new and different perspectives, understand himself and the world, and assess the situations and events he will face. Reading is a source of information that can be found online through books, periodicals, newspapers, magazines, newspapers, and a variety of other sources. Consequently, it encourages reading interests and academic success (Alsaeedi et al., 2021). A good reading habit is expected of university students because it is one of the requirements for both social and personal development. Understanding university students' reading preferences is essential for success in all academic areas and their continued personal growth (Bharuthram, 2017). As digital access and reading seem to become increasingly important in the current era for any country to achieve democracy, being digitally illiterate in developing African countries will lead to "ignorance, ignorance leads to deprivation, and deprivation leads to disintegration (Loan, 2009)." Digital literacy abilities are also necessary for reading academically (Stoller & Nguyen, 2020).

Many limitations and challenges related to digital resources and access face African developing countries, which libraries and librarianship professionals can help overcome if addressed adequately.

The promotion of digital reading has always faced some difficulties, including the following:

1. a lack of digital devices to access content;
2. resistance to change on the part of libraries and individuals;
3. a lack of adequate infrastructures, such as internet access;
4. a dearth of local content on electronic platforms; and
5. a lack of user awareness (Bouaamri et al., 2022).

## OBJECTIVES OF THE STUDY

1. To determine a student's level of digital literacy.
2. To identify the reading preferences of students.
3. To ascertain whether students prefer reading from electronic or print sources.
4. To be aware of students' proficiency with digital applications and software.

## METHODOLOGY

The study's current research design is a quantitative approach. It was a structured questionnaire. The questionnaire for this survey study was distributed using the random sample methodology. There are nineteen questions in all. It is divided into three sections. The first segment contained sociodemographic information, the second section addressed digital literacy, and the third section addressed the pupils' reading habits—students from DMI-St. Eugene University's BSc Computer Science and BE Computer Engineering programs served as the sample for the current study. Two hundred eighteen pupils were given the questionnaire, and 115 of them submitted accurate answers. Using a questionnaire, primary data for this study was gathered. The process of utilizing secondary sources involves conducting a literature review. Digital technologies and procedures were employed to gather reading habits and digital literacy data. The pupils received the questionnaire over WhatsApp. Employed the Five Likert Scale for this investigation. The SPSS 28.0 application was used to evaluate the data after completing the data collection process.

## RESULTS

**Table 1: Socio demographic Details**

| Type | Division | Respondents (%) |
|---|---|---|
| Gender | Male | 83 (73.2%) |
|  | Female | 32 (27.8%) |
| Age Groups (In years) | 17-21 | 102 (88.7%) |
|  | 22-27 | 13 (11.3%) |
|  | Above 27 | 0 (0%) |
| Location | Urban | 74 (64.4%) |
|  | Semi-Urban | 29 (25.2%) |
|  | Rural | 12 (10.4%) |
| Course | BSc Computer Science | 60 (52.2%) |
|  | BE Computer Science | 55 (47.8%) |
| **Total** | | **115** | **100** |

The demographic distribution of the respondents is shown in Table 1. 73.2% of respondents are men, compared to 27.8% of women, having a more significant percentage of men; In terms of age, 82.7% of respondents are between the ages of 17 to 21, 11.3% are between ages of 22 to 27, and 0% are older than 27; Geo location data reveals that 10.4% of respondents are rural, 25.2% are semi-urban, and 64.4% are urban; 47.8% of participants in the current school of study are BE computer science, while 52.2% are B.Sc. computer science.

**Table 2: Sources of information on emerging technology**

| Sources for knowing about new technologies | Respondents | Percentage (N=115) |
|---|---|---|
| Blogs | 52 | 45.2 |
| E-Commerce Apps | 35 | 30.4 |
| Email Lists | 12 | 10.4 |
| Friends | 87 | 75.7 |
| Magazines | 48 | 41.7 |
| Newspapers | 35 | 30.4 |
| Social Networks | 95 | 82.6 |
| Teachers | 22 | 19.1 |
| TVs | 74 | 64.4 |
| Websites | 73 | 63.5 |

Students use a variety of sources to learn about new technologies. According to table 2, social networks are the primary source for 82.6% of respondents, while friends account for 75.7% of respondents.

**Table 3: Self-rating of digital literacy skills**

| Digital literacy skills | Mean | SD | T-Value | P-Value | Result |
|---|---|---|---|---|---|
| Typing skill | 3.9 | 0.94 | -23.1 | < .00001 | significant |
| Web search skill | 4.21 | 0.83 | -32.2 | < .00001 | significant |
| Computer literacy | 4.19 | 0.62 | -44.2 | < .00001 | significant |
| Internet literacy | 4.08 | 0.81 | -31.4 | < .00001 | significant |
| Digital literacy | 4 | 0.76 | -32.71 | < .00001 | significant |

*Scale: Very Good=5, Good=4, Acceptable=3, Poor=2, Very Poor=1, SD= Standard Deviation*

Out of the five results, all five were statistically significant; as shown in table 3, statistical significance was discovered for all five individuals' assessed confidence levels. Web search abilities were statistically significant in relation to individuals' reported high literacy levels (M=4.21; SD=0.83) followed by computer literacy (M=4.19; SD=0.62), Internet literacy (M=4.08; SD=0.81), digital Literacy (M=4; SD=0.76), and lowest level value is typing skills (M=3.9; SD=0.94).

**Table 4: Digital literacy skills**

| Digital literacy skills | Yes (%) | No (%) |
|---|---|---|
| Do you know how to build and update websites? | 23 (20%) | 92 (80%) |
| Do you maintain a personal website or online portfolio? | 13 (11.3%) | 102 (88.7%) |
| Do you use any mobile applications to learn a foreign language? | 76 (66.1%) | 39 (33.9%) |
| Do you know how computer hardware works at its most basic level? | 106 (92.2%) | 9 (7.8%) |
| Are you proficient with keyboard shortcuts? | 97 (84.3%) | 18 (15.7%) |
| Are you familiar with social networking sites? | 75 (65.2%) | 40 (34.8%) |
| Do you use the computer for learning purposes? | 115 (100%) | 0 (0%) |

Table 4 data demonstrate the foundational level of digital literacy. Every student utilizes a computer to learn. 92.2% of participants can recognize the fundamental operations of computer hardware. 88.7% of students do not maintain websites, whereas 84.3% of students are efficient at keyboard shortcuts, and 80% of students are unaware of how to build web pages.

**Table 5: Frequency of using the digital environment**

| Utilization of the digital surroundings | Mean | SD | T-Value | P-Value | Result |
|---|---|---|---|---|---|
| Blog | 2.05 | 1.02 | -3.06 | < .002471 | significant |
| Database | 2.23 | 1.01 | -4.83 | < .00001 | significant |
| Electronic dictionary | 3.83 | 1.03 | -19.19 | < .00001 | significant |
| Graphic software | 2.43 | 1.14 | -5.46 | < .00001 | significant |
| Email | 3.69 | 1.07 | -16.97 | < .00001 | significant |
| Language App | 2.52 | 1.05 | -7.16 | < .00001 | significant |
| Spreadsheet | 2.76 | 1.02 | -9.7 | < .00001 | significant |
| Text chatting | 4.5 | 0.97 | -27.96 | < .00001 | significant |
| Video conferencing | 4.21 | 1 | -24.09 | < .00001 | significant |
| Voice chatting | 3.19 | 1.11 | -11.81 | < .00001 | significant |
| Word processor | 3.96 | 0.93 | -24.17 | < .00001 | significant |
| World Wide Web | 4.76 | 0.76 | -44.34 | < .00001 | significant |

*Scale Used: Very frequently=5, frequently=4, occasionally=3, rarely=2, never=1, SD= Standard Deviation*

As shown in table 5, the majority of participant's low level use of blogs (M = 2.05, SD=1.02), database (M = 2.23, SD=1.01), graphic software (M = 2.43, SD=1.14), language app (M = 2.52, SD=1.05) and Spreadsheet (M = 2.76, SD = 1.02). Medium levels of using digital environments are voice chatting (M = 3.19, SD=1.11), Email (M = 3.69, SD=1.07), e-dictionary (M = 3.83, SD=1.03), word processor (M

**Table 6: Self-rated competencies for digital applications**

| Digital application Skills | Mean | SD | T-Value | P-Value | Result |
|---|---|---|---|---|---|
| Word processing applications (e.g., MS Word) | 3.3 | 1.07 | -13.6 | < .00001 | significant |
| Spreadsheet applications (e.g., MS Excel) | 2.99 | 0.93 | -13.73 | < .00001 | significant |
| Database applications (e.g., MS Access) | 2.73 | 0.78 | -14.24 | < .00001 | significant |
| Presentation applications (e.g., MS PowerPoint) | 2.96 | 1.02 | -11.67 | < .00001 | significant |
| Communication applications (e.g., Skype) | 2.85 | 1.19 | -8.13 | < .00001 | significant |
| Learning management systems (e.g., Moodle) | 2.39 | 1.02 | -6.21 | < .00001 | significant |
| Social networking applications (e.g., Facebook) | 3.62 | 1.18 | -13.85 | < .00001 | significant |
| Web design applications (e.g., Dreamweaver) | 2.23 | 1 | -5 | < .00001 | significant |
| Web search engines (e.g., Google) | 3.51 | 1.22 | -12.22 | < .00001 | significant |
| File sharing sites (e.g., Dropbox) | 3.2 | 1.11 | -11.87 | < .00001 | significant |

*Scale Used: Excellent=5, Good=4, Average=3, Poor=2, Very Poor=1, SD= Standard Deviation*

= 3.96, SD=0.93). Findings revealed that participants rated their high levels of use on the World Wide Web (M = 4.76, SD = 0.76) followed by text chatting (M = 4.5, SD = 0.97), and video conferencing (M = 4.21, SD =1). According to further analyses of the data, there is a statistically significant difference.

Participants' ratings of their competence in digital application skills are shown in table 6. Overall, the results show that students' digital application skills are at a moderate level. Despite the fact that some skills, like low level understanding such as web designing application (M = 2.23, SD = 1) and learning management system (M = 2.39, SD = 1.02) (Ukwoma et al., 2016; Reddy et al., 2020). Additional data analysis showed that there was a statistically significant distinction.

As shown in Table 7, nearly all participants had prior experience using digital devices and felt comfortable doing so. Findings revealed that participants rated their confidence levels with those who like using technology, feel at ease utilizing digital devices, and believe it is vital for them to increase their digital fluency (M = 4.92, SD = 0.5). An additional examination of the data indicated the existence of a statistically significant difference.

**Table 7: Self-rated digital devices usage**

| Use of digital devices | Mean | SD | T-Value | P-Value | Result |
| --- | --- | --- | --- | --- | --- |
| Like using technology | 4.92 | 0.5 | -65.4 | < .00001 | significant |
| Feel at ease utilizing digital devices | 4.92 | 0.5 | -65.4 | < .00001 | significant |
| Aware of various types of digital devices | 4.55 | 0.81 | -37.87 | < .00001 | significant |
| Familiar with different digital devices | 4.27 | 0.92 | -28.14 | < .00001 | significant |
| In the context of digital technologies, eager to know more | 4.82 | 0.61 | -56.47 | < .00001 | significant |
| Believe that in terms of using digital tools, I lag behind my peers | 3 | 1.11 | -10.32 | < .00001 | significant |
| Believe that it is vital for me to increase my digital fluency | 4.92 | 0.5 | -65.4 | < .00001 | significant |
| Believe that utilizing digital tools and resources may improve my study | 4.74 | 0.65 | -51.78 | < .00001 | significant |
| Programs for language education ought to, in my opinion, incorporate instruction in technology-assisted language acquisition | 4.73 | 0.77 | -42.76 | < .00001 | significant |

*Scale used: Strongly Agree=5, Agree=4, Uncertain=3, Disagree=2, Strongly Disagree=1, SD=Standard Deviation*

Table 8: Preference for reading for pleasure

| Preference for reading for pleasure | Literary (books, drama, poetry) | Non-literary (magazines, journals, news, opinion, research) |
|---|---|---|
| Very much | 47 (40.9%) | 48 (41.7%) |
| Quite a bit | 58 (50.4%) | 48 (41.7%) |
| Moderately | 10 (8.7%) | 19 (16.6%) |
| Not much | 0 (0%) | 0 (0%) |
| Not at all | 0 (0%) | 0 (0%) |

Table 8 demonstrates the difference between readings for the enjoyment of literary or non-literary literature. Consequently, 40.9% of respondents in the above table liked literary books very much, whereas 41.7% liked non-literary books very much.

Table 9: Spending time each day reading

| Time spent reading | Respondents | Percentage |
|---|---|---|
| I do not read unless I have to | 29 | 25.2 |
| Less than 15 minutes | 0 | 0 |
| Between 15 minutes to 30 minutes | 21 | 18.3 |
| 30 minutes or more | 65 | 56.5 |

Table 9 indicates how much time each student spends consistently reading. More than half of the respondents (56.5%) read for even more than thirty minutes each day, while no one read for less than fifteen minutes.

Table 10: Electronic devices used for reading

| Digital gadgets did you utilize for reading | Respondents | Percentage (N=115) |
|---|---|---|
| Smartphone | 106 | 92.2 |
| Computer (Desktop) | 10 | 8.7 |
| Computer (Laptop) | 97 | 84.4 |
| Tablet | 45 | 39.1 |
| E-book Reader (e.g., Kindle, Nook) | 0 | 0 |
| Television | 23 | 20 |
| Portable Media Player (e.g., iPod) | 12 | 10.4 |
| Printed resources | 13 | 11.3 |

Note: allowed more than one alternative to be chosen so that the percentage is more than 100

Table 10 describes the electronic equipment used for reading. 92.2% of respondents used mobile phones, 84.4% used laptops, and no one used e-book readers. Additionally, only 8.7% of people are on a desktop (Chaudhry & Al-Adwani, 2019; Das et al., 2019).

**Table 11: Recommendations for books from**

| Reading recommendations received from | Respondents | Percentage (N=115) |
|---|---|---|
| A librarian or library, including library websites | 82 | 71.3 |
| Social media channels | 96 | 83.5 |
| Bookstores & employees (online & offline) | 43 | 37.4 |
| family, friends, and coworkers | 43 | 37.4 |
| Literary influences | 23 | 20 |
| Literary associations or book clubs | 0 | 0 |
| Reviews and news | 49 | 42.6 |
| A website that recommends books, like Amazon | 10 | 8.7 |

*Note: allowed more than one alternative to be chosen so that the percentage is more than 100*

Table 11 shows the majority of book recommendations (83.5%) came from social media platforms, followed by 71.3% from their university's library and websites, and none from literary organizations or book clubs.

**Table 12: Typically, setting read**

| Most reading settings | Respondents | Percentage (N=115) |
|---|---|---|
| In bed or a calm environment at home | 115 | 100 |
| In a vacation | 23 | 20 |
| Public spaces, like libraries | 33 | 28.7 |
| Traveling, using public transit, or waiting rooms | 22 | 19.1 |
| During brief quiet periods from other activities | 57 | 49.6 |

Table 12 looked at a reading-friendly setting for students. Out of 100 respondents, 100 chose their bed as their favorite reading spot, 49.6% chose a peaceful time away from other activities, and 28.7% said they only liked to read in public spaces like libraries.

**Table 13: Main motivations for reading enjoyment**

| Several good reasons to read | Respondents | Percentage (N=115) |
|---|---|---|
| It provides enjoyment and wisdom | 111 | 96.5 |
| Develops a sophisticated vocabulary | 105 | 91.3 |
| Better writer by reading | 98 | 85.2 |
| Gaining an understanding of particular matters | 85 | 73.9 |
| Developing creative thinking | 108 | 93.9 |
| Lowering anxiety | 44 | 38.3 |
| Aids in passing the time | 43 | 37.4 |
| Acquiring knowledge and learning new things | 104 | 90.4 |
| Enhancing imagination and creativity | 106 | 92.2 |
| Enhancing memory power | 91 | 79.1 |
| Enhancing focus and attention persistence | 76 | 66.1 |
| Knowledge of global perspectives on people, political commentary, geography, literature, and much more | 99 | 86.1 |
| Future educational excellence | 111 | 96.5 |

*Note: allowed more than one alternative to be chosen so that the percentage is more than 100*

Table 13 displays the reading objectives for students. 96.5% of respondents said they read for insight and pleasure, equally educational excellence. 93.9% of readers read to improve their creative thinking, followed by 92.2% of readers who read to increase imagination and creativity. Furthermore, 91.3% said they read to improve their language skills, while 90.4% of respondents said they read to gain knowledge and learn new things (Sun et al., 2020).

**Table 14: Reading Format preferred the most**

| Most preferred reading format | Respondents | Percentage |
|---|---|---|
| Electronic | 80 | 69.6 |
| Print | 35 | 30.4 |

Table 14 displays the preferred reading format among students. 30.4% of respondents prefer reading print, compared to 69.6% who prefer electronic reading (Alamri, 2019).

Table 15 illustrates an appropriate layout for reading books in various locations. In bed reading, respondents (57.4%) mentioned electronic, for 73.9% of respondents, electronic reading is enjoyable and has recreational value, 57.4% of respondents emphasized that using electronic resources during travel is more comfortable, 66.1% indicated that the most accessible format for sharing with others is electronic, The optimum method for obtaining and managing the extensive collection of books is electronic (91.3%), and 92.2% of people believe that quick access to new information is available digitally. Overall, most students believed that electronics were preferable to printed materials in this study

Table 15: Electronic or print books are more suited to you for the following

| Electronic or print books | Print | Electronic |
|---|---|---|
| Reading in bed | 49 (42.6%) | 66 (57.4%) |
| Reading pleasure / recreational value | 30 (26.1%) | 85 (73.9%) |
| Travel/commute reading | 49 (42.6%) | 66 (57.4%) |
| Sharing with people | 39 (33.9%) | 76 (66.1%) |
| Accessing and maintaining a wide collection | 10 (8.7%) | 105 (91.3%) |
| Quick access to new material | 9 (7.8%) | 106 (92.2%) |

## CONCLUSION

The popularity of digital learning environments among university students has dramatically increased. In the current environment, digital literacy and digital-based reading are particularly crucial. Various print and non-print resources are available in libraries. This study showed that digital literacy could be acquired everywhere, in urban and rural areas. Digital literacy skills are crucial for evaluating the materials, reading habits are vital for in-depth understanding, and web search ability is one of them found in this study. Findings also revealed that students frequently prefer using platforms like the world wide web and online chatting services for digital activity. Digital literacy serves as an anti-virus since misinformation and disinformation are spreading like a virus everywhere. Students are open to reading from any source, whether print or digital. Although, pupils are very comfortable reading from digital media. However, due to plenty of digital resources, they are moving towards electronic text instead of printed one, and the smartphone is the most preferred (Chaudhry & Al-Adwani, 2019; Das et al., 2019).

The purpose of reading may be different based on the individual. However, the common motive is to fetch information sometimes for pleasure, critical thinking, creativity, or to add knowledge to an idea. The digital environment has given this freedom to its reader to access information from varied sources at one go, so the respondents of this study steadily prefer to grab information mostly for pleasure or creative ideas during their bedtime. According to research findings, digital literacy should be included in the disciplines required to be taught at the university level. Libraries for students should subscribe to more electronic resources. Universities must enhance their pupils' digital literacy skills through workshops or training programs to address challenges faced while accessing digital platforms. Library professionals must start literary groups or clubs to inspire their student population to use digital devices not just for pleasure but for research also and educate them to access information from authentic sources (Breakstone et al., 2018). Libraries may also acquire e-reading devices to provide students easy access to pleasure reading.

**Acknowledgment**

We appreciate the assistance of Mr. Vinoth Arockiasamy, HoD, Department of Computer Science at DMI-St. Eugene University, in gathering the information needed for this study.